\documentclass[%
 aip,
 amsmath,amssymb,
 reprint,%
]{revtex4-1}

\usepackage{graphicx}
\usepackage{dcolumn}
\usepackage{bm}

\usepackage[utf8]{inputenc}
\usepackage[T1]{fontenc}
\usepackage{mathptmx}
\usepackage{etoolbox}
\usepackage{hyperref}
\usepackage{cleveref}

\makeatletter
\def\@email#1#2{%
 \endgroup
 \patchcmd{\titleblock@produce}
  {\frontmatter@RRAPformat}
  {\frontmatter@RRAPformat{\produce@RRAP{*#1\href{mailto:#2}{#2}}}\frontmatter@RRAPformat}
  {}{}
}%
\makeatother
\begin{document}


\title[Acoustic Flow Control using Reinforcement Learning]{Acoustics-based Active Control of Unsteady Flow Dynamics using Reinforcement Learning Driven Synthetic Jets}

\author{Siddharth.Rout}%
 \email{siddharth.rout@ubc.ca.}
\author{Khai Phan}
\affiliation{ 
   University of British Columbia,\\
   Vancouver, BC V6T 1Z4, Canada
}%

\author{Chao-An Lin}
\affiliation{ 
   National Tsing Hua University,\\ 
   Hsinchu, 30013, Taiwan
}

\date{\today}

\begin{abstract}
Flow generated noise are caused shear flows and, hence, they can be used as feedback to control the flow. Existing flow control uses state variables like velocity, pressure, or vorticity, none use acoustic observables as the primary control signal. It is tough to model a classical control algorithm using sound level but data-driven approaches are not as do not have to explicitly model the physics. We present an acoustics-driven framework for active control of unsteady wake dynamics behind a circular cylinder, in which sound is used as the primary feedback signal for flow regulation. The approach integrates deep reinforcement learning (DRL) with synthetic jet actuation, using acoustic measurements acquired from a downstream hydrophone array to inform control decisions in real time. Unlike conventional flow control strategies that rely on velocity or pressure field sensing, the proposed method establishes a direct link between far-field acoustic emissions and near-field actuation. Within this formulation, the DRL agent learns control policies that exploit acoustic signatures of vortex shedding to modulate synthetic jet actuation on the cylinder surface. The resulting control suppresses coherent wake structures and mitigates flow-induced disturbances. Quantitative results show reductions of up to 9.5\% in radiated noise and 23.8\% in drag under the tested conditions, accompanied by a marked attenuation of wake oscillations, for a DFG 2D benchmark flow with Reynolds number 100. These findings demonstrate that acoustic sensing alone can provide sufficient information for effective closed-loop flow control and highlight its potential as a non-intrusive feedback modality for coupled aerodynamic and aeroacoustic optimization in bluff-body flows. The codes for the algorithm can be found here: \url{https://github.com/Siddharth-Rout/FlowControlDRL}.
\end{abstract}

\maketitle

\section{Introduction}
\label{intro}
Flow control has always been one of the most anticipated engineering problems due to its ubiquitous applicability. From the suppression of flow oscillation in open cavities \cite{CATTAFESTA2008479} to the construction of hybrid rocket motors \cite{TAN2023238}, flow control has been used as an indicator of how technology has developed to counter the stochasticity of nature. Throughout the last few decades, the number of paradigms for flow control has been increasing more and more. Applications of flow control to air vehicle systems, including fixed wing airfoils, turbomachinery, combustion, aeroacoustics, vehicle propulsion integration, and rotorcraft. Flow control methods can be categorized into Active Flow Control (AFC) and Passive Flow Control (PFC). There are many innovative applications established in various industries\cite{Joslin_Miller_2009, wang_feng_2018}. PFC have a constant control law that is consistent with time and do not get any feedback on how well the controller performs, such as having changes to aerodynamic shapes or textures. The passive methods include Gurney flap, vortex generator, bump, cavity, roughness, small disturbance, bleed, splitter plate, polymer, and biomimetic techniques\cite{bio_flow_control, 2023CmEng...2...30O}. Some examples are leading-edge serrations, riblets, corrugated airfoils and lubricated skins. Most of these are widely implemented in aircrafts to delay flow separation and increase lift to drag ratio. Winglets are nowadays found to be used to reduce tip vortex formation to reduce drag. However, these control strategies are limited as the control can not be manipulated temporally based on feedback or requirements. For instance, what if the PFCs act adversely? So, AFC is a good way out as it can take in feedback from the state and actuate the controller intelligently. The active methods include oscillation and flow perturbation, acoustic excitation, jet, synthetic jet, plasma actuator, and Lorentz force. Many interesting AFCs developed in past decades. For example, installation of the synthetic jet to change the vortex shedding pattern\cite{FENG201414}; utilizing wavelength actuators to attenuate turbulence\cite{BhattGreg}; and studying the effects of acoustic excitation on vortex shedding \cite{FUJISAWA2004159}. Among those numerous methods, the application of blowing-suction of velocity jets stands out as one of the most practical and widely recognized, evidenced by NASA’s experiment on the Boeing 757 with jet actuators incorporated in the vertical stabilizer to reduce drag and improve the overall performance of the plane \cite{JohnCLin, NASA_2023}. 

Why specifically controlling flow past a cylinder? Flow-induced forces play determinant roles in the life and safety of structures as well. Oscillations in the flow cause fatigue, enhance defects, aeroelastic flutters, and decrease the factor of safety of structures. Falling of the famous Tacoma Narrows Bridge is a popular case of structural failure due to similar causes\cite{ARIOLI2015901}. Tall buildings like Taipei 101, Burj Khalifa etc. have to be designed to be able to face fast winds\cite{taipei101, GU2022103681}. 
Passive techniques developed in the past few decades are still very promising due to the ease of utilization in industry\cite{OWEN2001597, BAEK2009848, Law2017WakeSM}. AFCs past cylinders are excellent toy problems to demonstrate concepts. The oppositely placed suction and blowing around cylinder became popular active control strategy in 2000s\cite{KimChoi,Dong,WangTang}. To improve the quality nonlinear control algorithms\cite{mao_blackburn_sherwin_2015} and eigensystem-realization based reduced order model for suppression of wakes\cite{yao_jaiman_2017} are popular. However, these deterministic control algorithms like proportional–integral–derivative(PID) controllers often require approximation of state space and calculation of the transfer function to actuate AFCs is expensive yet inaccurate and non-generalizable as the transfer function is case dependant. Hence, data-driven model-free methods like reinforcement learning (RL) for AFC are well appreciated as they are generalizable.

In the past few years, there has been a surge in Deep Reinforcement Learning (DRL) based flow control techniques\cite{WangMei,yousif_zhang_yu_yang_zhou_lim_2023,li_zhang_2022,Dfan}. This work is hence based on a DRL algorithm to be able to extend the work to realistic cases like flow past marine vehicles. A few recent attempts to utilize DRL in AFC include controlling two synthetic jets of blowing/suction \cite{rabault_2019} and implementing adjoint-based partial derivative equation augmentation to DRL to solve flow simulation more efficiently\cite{liu2024adjoint}. Both showed great results in controling wakes. In order to reduce vortex shedding, the two aforementioned papers both tried to decrease the drag coefficient in the simulation. However, no work is registered in which acoustic-based flow control model is used to reduce wakes formation and control flow dynamics. Unlike many existing DRL-based active flow control studies that rely primarily on velocity-field observations or force coefficients as feedback variables, the present work investigates acoustically informed pressure-feedback measurements as the principal control signal for adaptive wake stabilization. The study therefore focuses on the interaction between wake-induced hydrodynamic pressure fluctuations, aeroacoustic response, and DRL-guided synthetic jet actuation.    

Deep Q-Nework (DQN) is a branch of DRL that involves calculating the Q values of each step the model takes and, much like other ML algorithms, learning to maximize the returned reward from the environment. DQN seems to function well in more abstract tasks and therefore is widely used in robotics and has achieved great success in this field. For instance, Fernandez-Fernandez et al. study the application of DQN to human-like sketching performed by robots\cite{FERNANDEZFERNANDEZ202357}. In the context of of controls and optimization in fluid dynamics morphing airfoils and shape optimization using DQN is also evident\cite{rout2022airfoil}. Despite the level of sophistication of the model, it is not very widely used in AFC modelling. In addition, manipulation of other properties apart from drag calculation is not common in the overall scientific conversation in AFC. According to Klapwijk et al., turbulence in the fluid flows is the source of sound generation in the system. The article explores noise levels when the turbulent flow is increased\cite{KLAPWIJK2022111246}. Interestingly, it also claims that the noise generation mechanism is difficult to understand. In this work, it is shown that by controling vortex generated noise using deep Q-learning drag and wake amplitudes could be controlled.  

In the nineteenth century, wakes are identified as the major source of noise in the flow past an object. There lies the concept behind this work and it is theoretically supported. Strouhal in 1878 and Kohlrausch in 1881 independently found out about a faint sound originating from vortices, to which the latter described as 'reibungstone' \cite{strouhal, stefanini, Rout23}. Sir James Lighthill, in 1950s, discovered the theoretical connection between fluid flow and acoustics from the conservation laws to derive the wave equation for acoustics, called the theory Lighthill's Acoustic Analogy \cite{Lighthill}. Sir Lighthill creates an experiment assuming a patch of turbulent flowing fluid surrounded by a large domain of surrounding stationary fluid. Let turbulent flow produce noise; however, the noise would transmit to the surrounding fluid at rest. By analysing and comparing the terms in the conservation equations for stationary fluid, the resulting equation could be written as a forced bidirectional wave equation. It becomes clear that the forcing term or the source term in the wave equation is what generates noise in flow. Here is the derived wave equation in Einstein's notations,
\begin{equation}
    \frac{\partial^2 \rho}{\partial^2 t}
   - c_o^2 {\nabla^2\rho} = \frac{\partial^2 {T_{ij}}}{{\partial x_i}{\partial x_j}}
\end{equation}
\begin{equation}
    {T_{ij}} = {\rho}{v_iv_j}- {\sigma_{ij}} + (p - c_o^2\rho){\delta_{ij}} 
\end{equation}
where \textbf{T} is called Lighthill's turbulence stress tensor and has three components or three sources for noise generation. \({\rho}{v_iv_j}\) is the convection of momentum fluctuation, \( \mathbf{\sigma}\) is the viscous stress tensor and \((p - c_o^2\rho)\delta_{ij}\) is the difference in exact pressure \(p\) and approximated thermodynamic pressure, \(c_o^2\rho\). \({\rho}\) is the density, \(c_o\) is the speed of sound, \(t\) is the time dimension, \(\textbf{\textit{x}}\) is the spatial dimension and \(\textbf{\textit{v}}\) is the velocity. This equation quantifies sound sources from flowing fluid where it takes care of thermodynamics jumps, turbulent fluctuations, and viscous dissipation. Now the fact is if the stress tensor is non-zero sound is formed and with the origin of wakes, vortex stretching starts into action and even with laminar vortex street sound is produced. This sound is often considered tonal while a broadband noise is generated in case of turbulent flows. Various further works have demonstrated this \cite{inoue_hatakeyama_2002,kumar,LIOW2006407}. 

In this paper, due to the limitation of computational resources, the simulations are conducted using incompressible flow. This is based on this simple assumption that incompressible flows still produce acoustic-like pressure fluctuations, though incompressible flow solvers assume constant density. This is because the Lighthill's stress tensor (\textbf{T}) is non-zero. The speed of sound in such cases tends to infinity. As such speed of sound being much slower than the speed of fluid, Mach number << 1, the nearfield observations are barely affected \cite{crighton1993computational,meister2002computational,Ask2003AnAA, layton2009lighthill}. We assume these pressure fluctuations as approximate acoustic noise. We essentially use a proxy acoustic metric derived from hydrodynamic pressure fluctuations. So, wakes make noise like pressure fluctuation and hence definitely if louder is the noise in a flow then stronger are the wake vorticities. Though this logic is very much valid but it has not been used for control or analysis rather flow states like pressure and velocity fields are considered as direct measurements and sometimes vorticity field is used as a direct measure of rotational energy in a flow. In this research, we aim to minimize the wake formation in the flow and hence lower the specially calculated effective sound pressure levels (SPLs) created by the vortices in the flow past a stationary circular cylinder using the flow-generated sound. Source of sound is a better and easier way to control vorticity in a flow as vorticity is the source of generated noise, hence being a more logical measure for such flow control problems. Furthermore, we will explore the effect of lowering SPLs on the oscillating drag experienced by the cylinder. As for the active control algorithm aspect of the research, DQN based reinforcement learning is used to control the blowing-suction of two synthetic jets on opposite ends of the cylinder perpendicular to flow.

This research is organized as follows. In order, sections 2,3, and 4 are dedicated to DQN-based control algorithm, introducing in detail the setup of the model used in the simulation, and the jets' actuation. Section 5 introduces SPL formulation and section 6 discusses the control strategies with experiments and results. Thence, section 7 concludes the work.

\section{Deep Q-Learning} 
Constructed on the Markov Decision Process in which the quality of action at a particular state is learned based on the reward due to the action, Q-Learning, has been a very popular early reinforcement learning\cite{watkins1989learning}. Conceptually, the action at a particular state is independent of the historical state following Markov probability. However, the rewards are learned from the cumulated score of rewards in an episode of the control process. The convergence of the optimality control problem using the Bellman Equation under stochastic updates was proved soon after\cite{Watkins1992}. The limitation of Q-Learning is the finite nature of the map between state to best action, which is called Q-table. Deep neural networks as excellent maps in Q-Learning replace the Q-table to get called Deep Q-Network (DQN) and the algorithm is called Deep Q-Learning. It is a breakthrough reinforcement learning algorithm since it has been able to match human-level control of console to Atari games\cite{dqn}. DQN allows mapping to conditional non-linear control algorithms. The second benefit is the possibility of being trained in an infinite and continuous environment state space.  

\begin{figure}[h]
\centering
\includegraphics[width=0.5\textwidth]{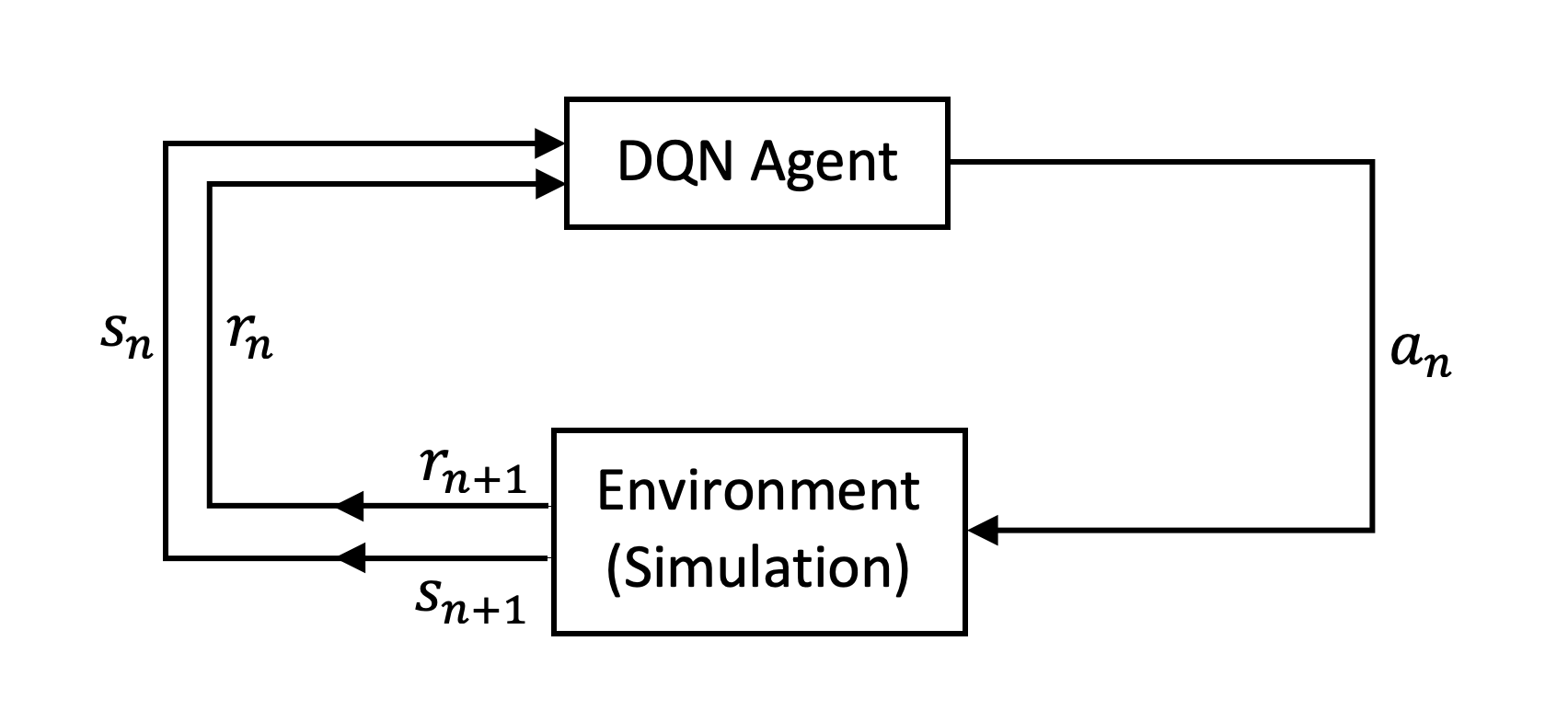}
\caption{Interaction between DQN agent and Environment in Markov Decision Process.}
\label{fig:RL_flow}
\end{figure}

The DQN agent learns to estimate and optimize the Q-values, which represent the expected rewards of an action taken in a particular state. This process is carried out on the neural network that examines all possible actions that result in the Q-values as outputs. The Bellman equation is then used to bridge the gap between predicted Q-values and target Q-values. 
\begin{equation}
    \mathbf{V}(\mathbf{s}) = max (\mathbf{R}(\mathbf{s},\mathbf{a}) + \gamma \mathbf{V}(\mathbf{s'}))
\end{equation}
$\mathbf{V}$ is Q-value; $\mathbf{R}$ is the reward of action $\mathbf{a}$ in state $\mathbf{s}$; $\gamma$ is the discount, representing the importance of immediate and future rewards; and $\mathbf{s'}$ is the following state. The algorithms can have a deterministic policy or a stochastic policy. A deterministic policy maps each action to a specific state, while a stochastic policy operates upon the probability distribution of the actions.

\begin{figure*}[t]
\centering
\includegraphics[width=1\textwidth]{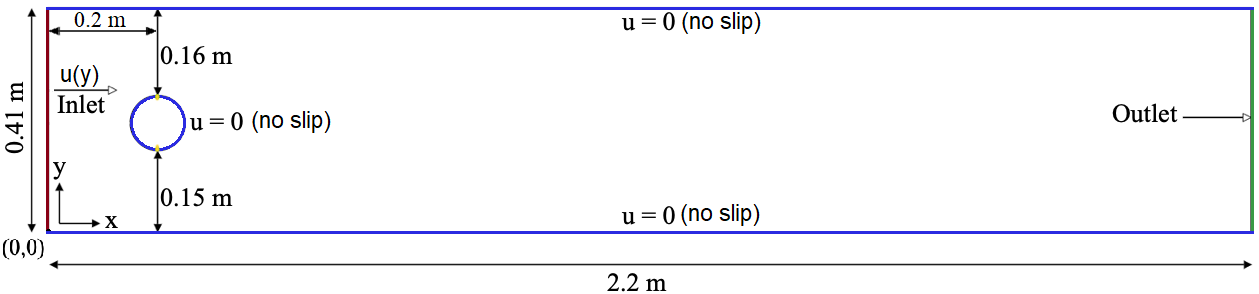}
\caption{Diagram of flow past cylinder setup, the jets are positioned at the yellow dots shown on the top and bottom surface of the cylinder.}
\label{fig:schematicdiagram}
\end{figure*}

The standard DQN algorithm is used as the reinforcement learning framework. However, certain modifications are made to make the DQN more compatible with the fluid mechanics nature of the project. Although most DQN algorithms are inherently well suited for deterministic problems due to limited and discrete state space in episodic environments. However, fluid flow field is a continuous state space and the algorithm is introduced to stochasticity with a random exploration strategy, initialization and sampling of mini-batches to interact with the stochastic environment in the flow. However, this can cause great instabilities and might prevent the algorithm from converging to a desired state. Therefore, we divide the simulation into a stochastic exploration stage and a deterministic testing stage.   

Regarding the construction of the neural network of the DQN Agent, four fully connected layers are made between the input and output layers. The input layer takes in the calculated SPL and the output layer gives out the two jet velocities. Each layer consists of 50 neural nodes which let 7902 learnable parameters. ReLU is the nonlinear activation function used at each node, which essentially acts as switch, hence making DQN a multilayer nonlinear switch. Adam optimizer\cite{kingma2014adam} in PyTorch library is applied to maximize rewards returned in each episode. More information about the utilization of the DQN agent in different tasks will be provided in the following section. The purpose of this task is to minimize the sound pressure levels (SPL) created by the wakes past the cylinder. There are multiple specific setup details in order to obtain the SPL reduction. 

\section{Problem introduction and setup}
The computational setup is built on DOLFINx, which is a high-performance solver of partial differential equations written in C++ for backend integration with legacy FEniCSx(version 2019.1.0) and python for interface\cite{DLFNx, DLFNx2}. The project allows the use of the standard benchmark case "Flow past a cylinder (DFG 2D-3 benchmark)”, as a simulation framework to further develop the research based on \cite{Langtangen2016, Schäfer1996, basecase}. The setup includes a horizontal rectangle with a height of 0.41m and a length of 2.2m, and the bottom left corner of the rectangle is at coordinate (0, 0). The obstacle is a circular-based cylinder with a radius of 0.05m centered at coordinate (0.2, 0.2). As the flow develops its oscillation, though laminar, the obstacle will experience a drag force, $C_{D}$, which can be determined using the formula:

\begin{equation}
    C_{D} = \frac{2}{\rho U^{2}_{mean}L}\int_{\partial \Omega_{S} }\{ \rho \nu \mathbf{n.}\mathbf{\nabla} u_{t_s} (t) {n_y} - p(t){n_x} \} ds,    
\end{equation}

where $u_{t_s}$ is the tangential velocity component at the interface of the obstacle $\partial \Omega_{S}$, defined as 
\begin{equation}
    u_{t_s} = \mathbf{u}.({n_y}, -{n_x}),
\end{equation} 
$\mathbf{n}$ is the normal unit vector at the surface, $n_x$ and $n_y$ are the x-component and y-components of normal vector, $U_{mean}$ the average inflow velocity, $\rho$ the fluid density, $\nu$ kinematic viscosity and $L$ the characteristic length of the cylinder, which is the diameter in this case.

The uniformly separated measurements around the cylinder can be used to determine the drag coefficient by summation of discrete measurements as approximate integration. For further details about the dimensions of the setup, refer to \Cref{fig:schematicdiagram}. 

Inflow is actuated from the left wall (near the cylinder) with a parabola shape according to the following formula for velocity: 
\begin{equation}
   u(y) = \frac{4Uy(0.41-y)}{0.41^{2}} 
\end{equation}
$y$ is the y-coordinates, and $U$ is $1.5 ~ ms^{-1}$ in this scenario, instead of the sinusoidal profile in the test problem as provided by Turek \cite{Schäfer1996, basecase}. So, the mean velocity 
\begin{equation}
    \overline{U} = \frac{2}{3}U = 1.0 ~ ms^{-1}.
\end{equation}
Furthermore, the outflow is the rightmost wall. The upper, lower, and obstacle walls all have a non-slip condition ($u=0$) as presented in \Cref{fig:schematicdiagram}. The kinematic viscosity is 0.001 and diameter ($D$) of the cylinder is $0.1$. So, the Reynolds number,
\begin{equation}
   Re = \frac{\overline{U}D}{\nu} = 100. 
\end{equation}

\section{Jets Configuration}
Two jets with blowing and suction control are used to manipulate the flow. The first jet (referred to as \textbf{Jet 1}) is at the top of the circular base of the cylinder (at coordinate ($0.2$, $0.25$)), and the second jet (\textbf{Jet 2}) is at the bottom (at coordinate ($0.2$, $0.15$)). The width ($w$) of the jets is small, at $0.25$ percent of the diameter, i.e. $2.5e{-4}~m$. The jets can perform blowing and suction independently, meaning blowing and suction can happen simultaneously. A reinforcement learning algorithm is applied to control the blowing and suction of the jets. More information about the execution of the simulation will be provided in the next section. 

\section{Feedback Formulation}

\begin{figure*}
\centering
\includegraphics[width=1\textwidth]{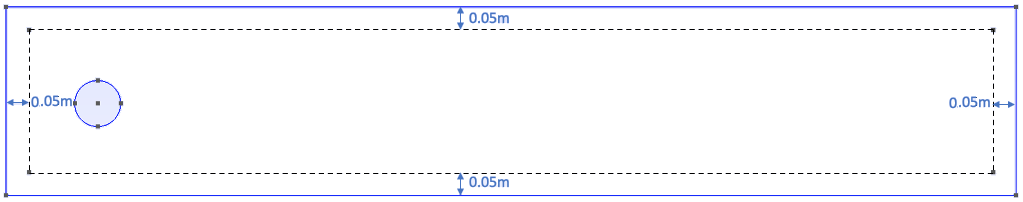}
\caption{Sensors positioned along the dashed line in 2D around the cylinder}
\label{fig:sensor_placement}
\end{figure*}
\begin{figure*}
\centering
\includegraphics[width=1\textwidth]{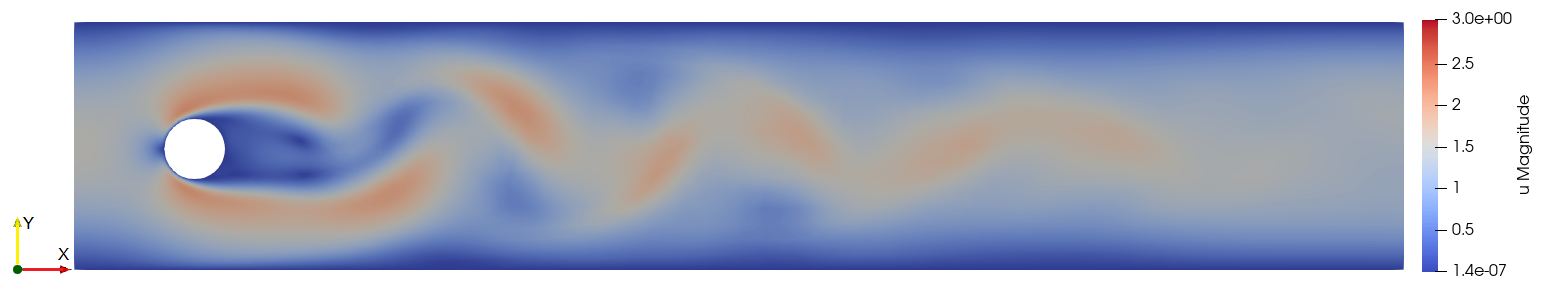}
\caption{Flow velocity field and visible vortices downstream to the cylinder, called vortex stream}
\label{fig:vortex_shedding}
\end{figure*}

To measure the SPLs, pressure is recorded from the pressure field provided in the simulation with surfaces of closely located sensors around the cylinder, 0.05m away from the container’s walls (refer to \Cref{fig:sensor_placement}). The sensor surfaces enclose the vortex street created by the flow and therefore can give a more accurate reflection of the varying pressure field along the vortex street and it helps us determine the static pressure level, which are essential for the calculation of the SPLs. We set 2000 sensor points horizontally and 500 sensor points vertically on each side. The pressure of each point at a particular time is then extracted from the pressure field created by the simulation. The upper and lower sensor surfaces are mainly concerned as they cover the length of the vortex street, which will be the source of most of the noise generation. These vortices are born from instabilities in the bottom and top regions of flow separation alternatively. Hence, having a distinction between the top horizontal sensor array and the bottom horizontal sensor array helps in understanding the vortex periodicity. Pressure values of every sensor point at each time step are recorded and passed through a function to convert to the relative $SPL_{i}$ for each sensor and effective SPL for the system $SPL_{eff}$ using the formulae below.
\begin{equation}
    SPL_{i} = 20\mathrm{log}\frac{|p_{i}-p_{avg}|_{rms}}{p_{avg}}
\end{equation}
\begin{equation}
    SPL_{eff} = 10\mathrm{log}\frac{\sum_{i=1}^{n} |p_{i}-p_{avg}|^{2}_{rms}}{p^{2}_{avg}} 
\end{equation}
With $p_{i}$ and $p_{avg}$ are the pressure value of each sensor and the average pressure value of all the sensor points at that time step averaged over previous 2000 time steps, that ensures to capture sufficient number of pressure oscillations to approximate the static pressure in the environment that is dynamic in nature. The SPLs at multiple sensors are then passed through another function to calculate the root-mean-square value, which is plotted to see the behaviour of the overall noise level in the environment.  

\begin{figure}[!t]
\centering
\includegraphics[width=0.48\textwidth]{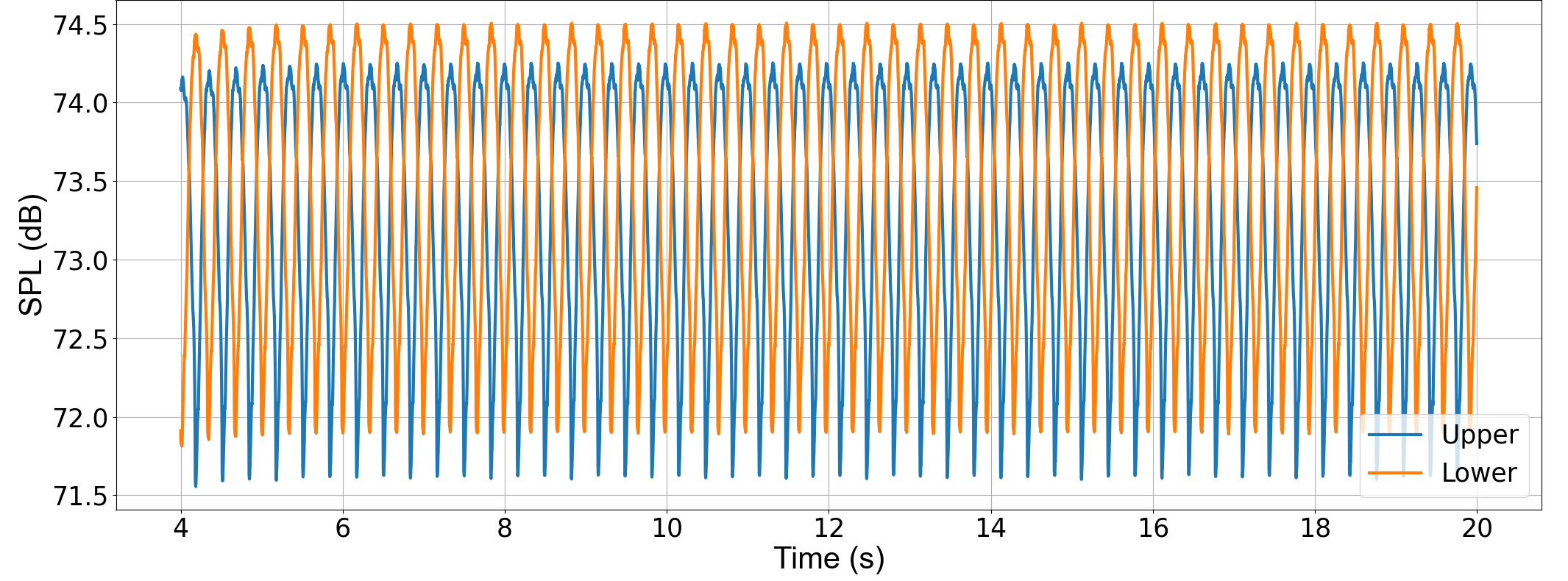}
\includegraphics[width=0.48\textwidth]{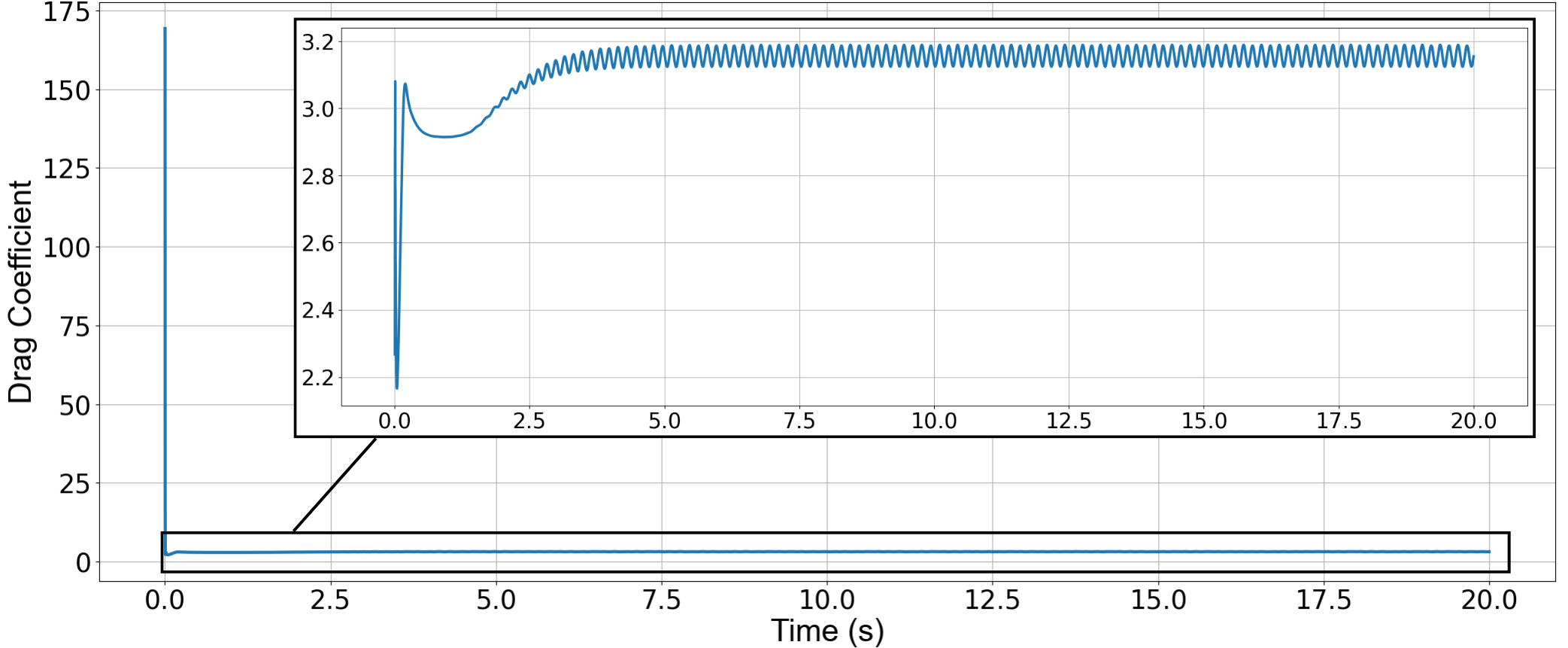}
\caption{a) SPL without jet intervention, b) Drag coefficient without jet intervention}
\label{fig:uncontrolled}
\end{figure}

\section{Experiments and Discussions}

The simulation runs for 20 seconds, from $t = 0$ to $t = 20$, to see the full behaviour of the SPL, as well as to allow the DQN algorithm ample time to learn and optimize. Each second has 500 time steps, resulting in 10,000 time steps to be solved overall. Regarding the recording process, we start by allowing the flow to develop and form the vortex street for the first 6 seconds. Then, when the oscillation stabilizes, the jets are let to intervene at $t = 6$. Due to computational limitations, the jet velocity values change every 50 time steps, which corresponds to a frequency of 10 Hz. A drastically rapid rate in the flow may over-influence the flow and alter it completely. Moreover, computational limitations of the simulation also play a part in this jet interjection.  
\subsection{Stage 1: Explore the optimal range of jet values}

\begin{figure}[!t]
\centering
\includegraphics[width=0.48\textwidth]{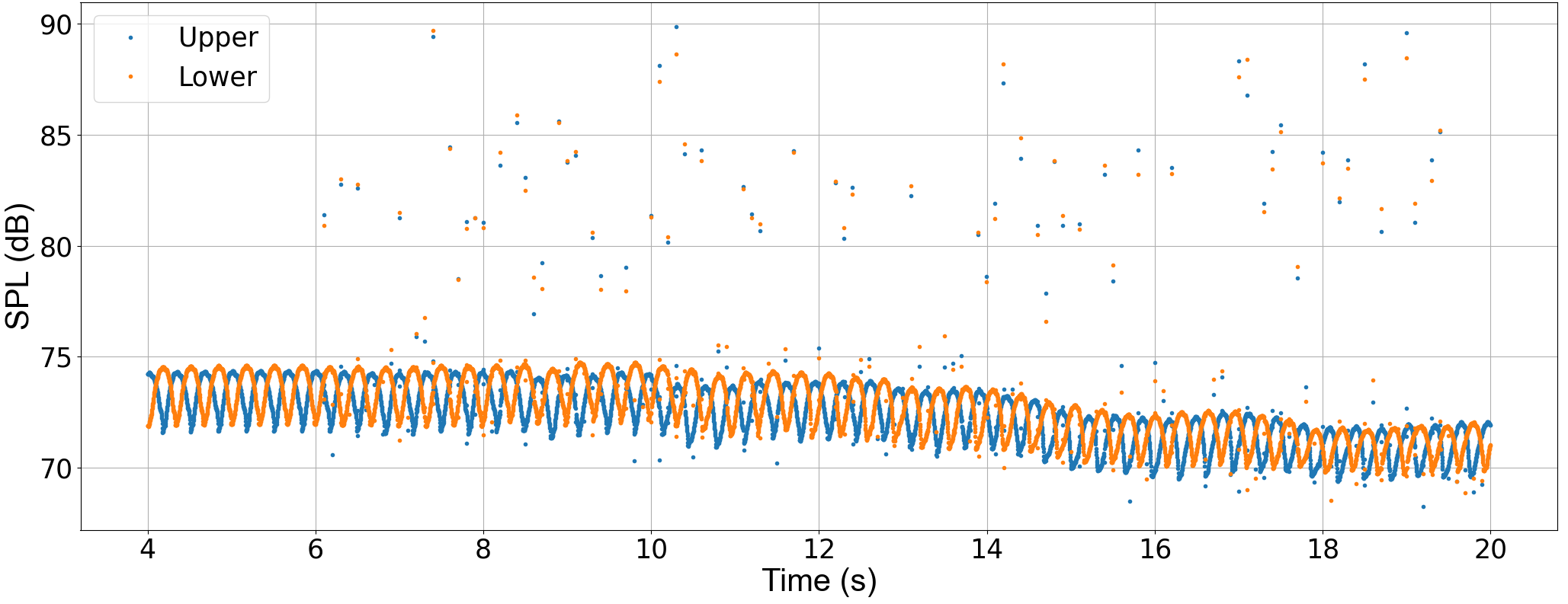}
\includegraphics[width=0.48\textwidth]{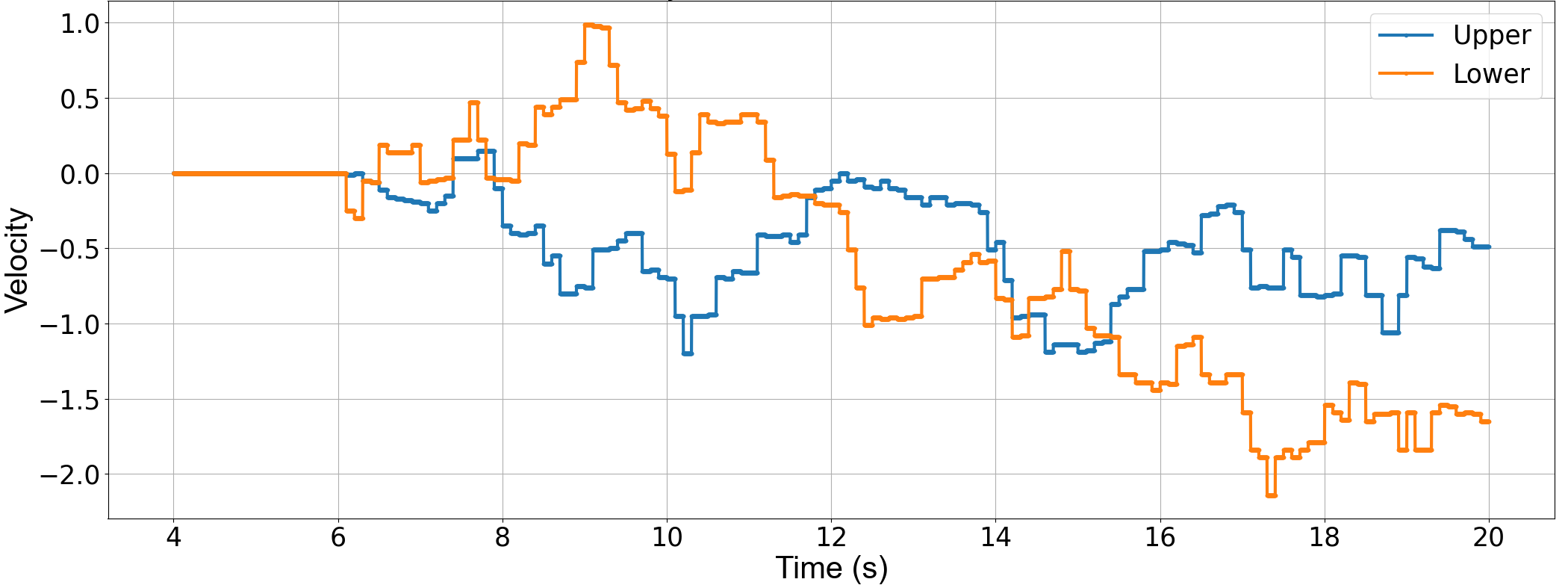}
\includegraphics[width=0.48\textwidth]{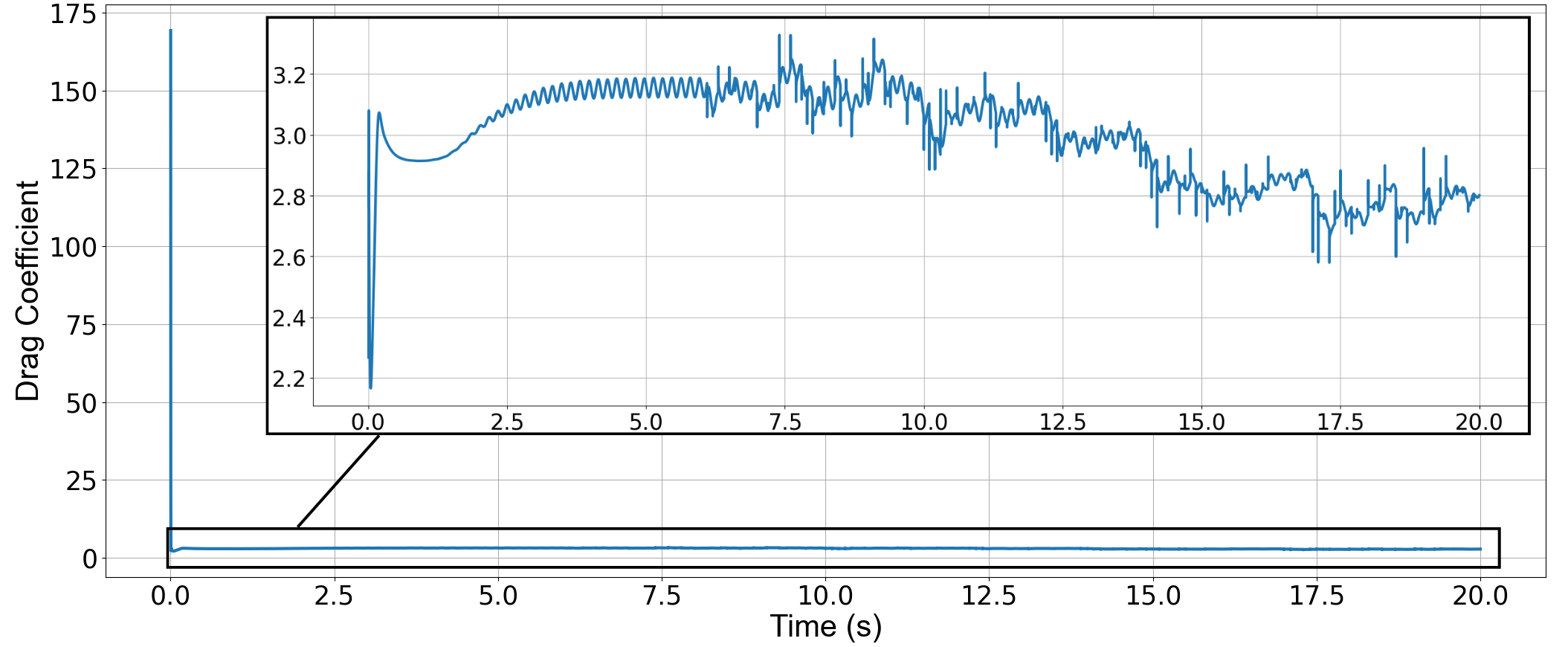}
\caption{After applying jets to reduce SPL using the strategy described in Stage 1 a) Overall SPL b) Velocity values of Jet 1 (Upper) and Jet 2 (Lower) c) Drag coefficient throughout the simulation}
\label{fig:stage_1}
\end{figure}

\begin{figure}[!t]
\centering
\includegraphics[width=0.48\textwidth]{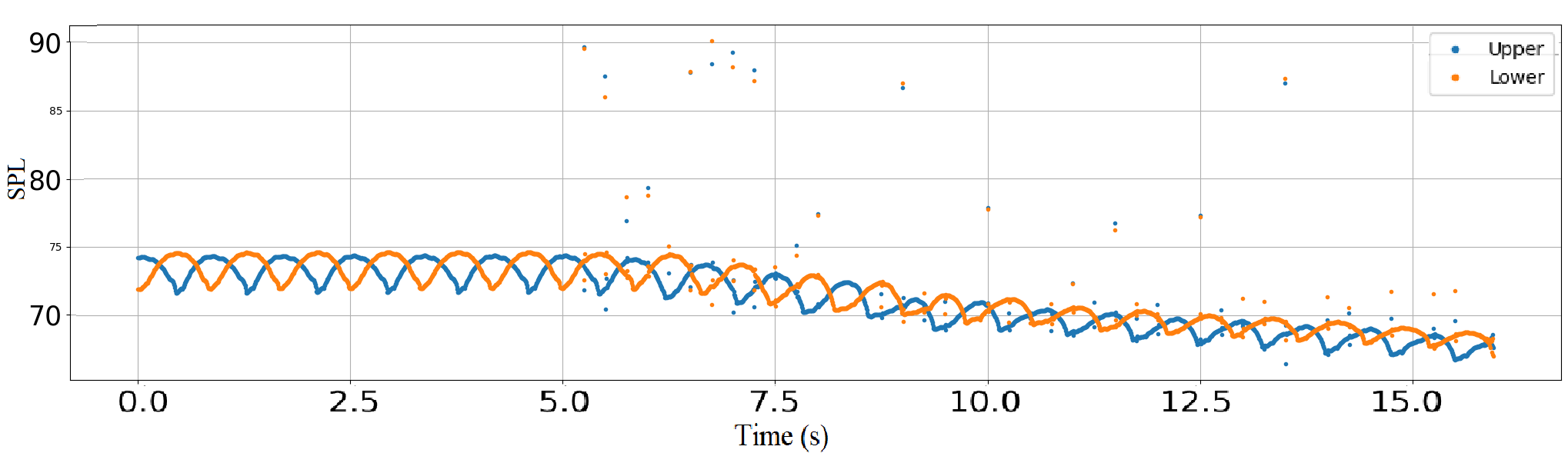}
\includegraphics[width=0.48\textwidth]{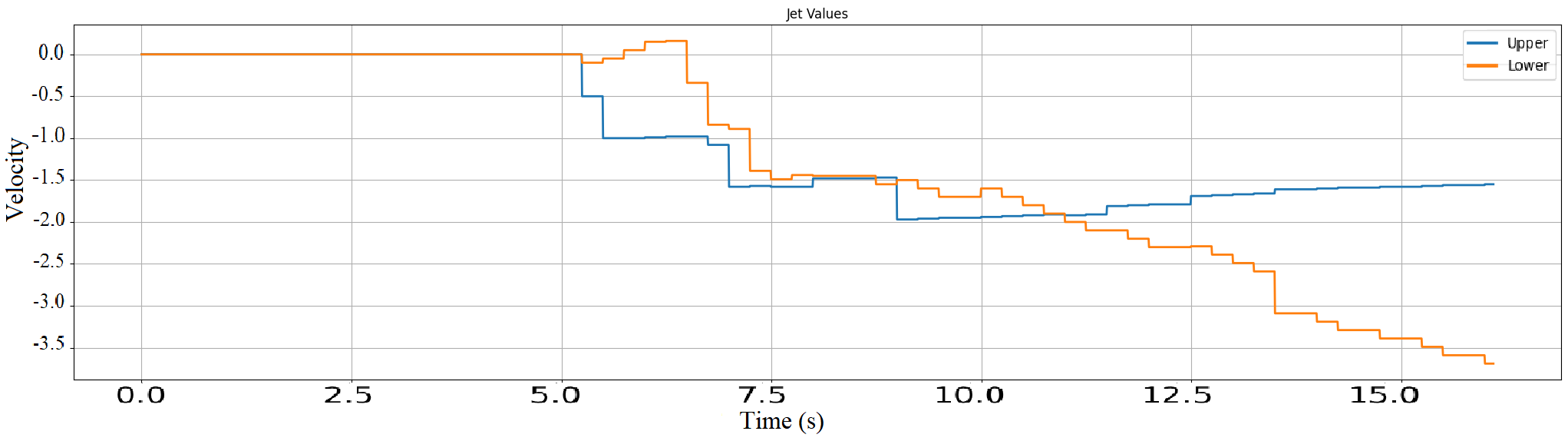}
\caption{a) After applying jets to reduce SPL using the strategy described in Stage 1 b) Velocity values of Jet 1 (Upper) and Jet 2 (Lower) with the second setting that encounters instability in simulation}
\label{fig:stage_1_test2}
\end{figure}

\subsubsection{Set-up}
It is challenging to determine what velocity value of the jet is able to influence the flow as we desired. Therefore, build-up strategy is used for the DQN to explore the value that can lower the SPLs. The velocity values of the jets are generated based on the values of the previous time step. The increments and decrements include ±0.01, ±0.05, ±0.1, ±0.5 and 0, so there shall be 81 combinations of actions for the DQN algorithm to manage. For example, if both jets initially have the values of 0.1, the algorithm will have the option to decrease Jet 1 by 0.01, 0.05, 0.1, 0.5 or keep it constant, and likewise for Jet 2. This results in a large number of actions for the DQN to explore, which can potentially hinder the ability to optimize the reward as the optimization problem needs more variables to be tuned and the gradient surface becomes rough and noisy, though the capability of optimization with more variables is better. Hence, fewer controlling variables are typically preferred. The velocity values are kept constant between the two interventions. The input of the DQN is a vector with a dimension of 1x4. The first two are for the SPLs of upper and lower sensor surfaces and the last two are the velocity values of Jet 1 and Jet 2. In this stage, we follow the model in \Cref{fig:RL_flow}, which allows the simulation to pass the data (reward, action, states) to the DQN at every time step. However, the jet velocity and therefore the action are kept constant every 50 time steps.       

The reward returned is determined by a function in the simulation. In this task, the simulation learns how to reduce approximately 3 - 5 dB and create a convergence in the process. A simulation runs without any jet intervention results in two oscillations. While the lower SPL peaks at roughly 74.5 dB, the upper SPL’s peak is about 0.25 dB lower. Moreover, the range of the SPL is from 71.7 to 74.5 dB (2.8 from maximum to minimum). Refer to \Cref{fig:uncontrolled} for further information. Meanwhile, the drag coefficient is also measured due to its popularity in wake control, so this measurement can be used as a means of verification for the noise reduction method. The drag experienced initially is large due to the direction of flow at the beginning. When the flow stabilizes, the plot suggests it has an oscillatory behaviour, with a maximum and a minimum at approximately 3.185 and 3.123, respectively.

The reward function is set as below:
\begin{table}[h]
\centering
\begin{tabular}{ |c|c|c|c| } 
 \hline
 SPL (dB) & Reward & SPL (dB) & Reward \\ 
 \hline
 \hline
 $>$ 74.5 & -10 & 73.0 - 73.5 & 0 \\ 
 \hline
 74.0 - 74.5 & -7.5 & 66.0 - 73.0 & 10 \\ 
 \hline
  73.5 - 74.0 & -5 & $<$ 66.0 & (*) \\ 
 \hline
\end{tabular}
\caption{Reward policy based on the returned states}
\end{table}

(*) -1 for every 0.4 dB below 66.\\

\subsubsection{Results}  
After applying the jet intervention and running the simulation to its completion, the result is plotted in \Cref{fig:stage_1}. From t = 4 to t = 6 on the x-axis, the SPL is stable because there is no intervention yet. However, from t = 6, there are many fluctuations in the SPL due to the exploration of the DQN algorithm. Many large values of jet values are chosen, which leads to observable discrepancies before the 8-second mark. The overall SPL in this time range is still relatively the same as when the jet is not turned on, but there is a small surge near t = 8, demonstrating the model has been able to achieve a jet pattern that can affect the SPL oscillations. 

From t = 8  to t = 10, the SPL continues to rise to a peak of 75 dB, then gradually falls over time. It is also evident that there are some interventions that can bring the SPL down, preventing it from increasing drastically. The act of exploration searches for optimal policy control by letting random actions to learn the optimality like mutation steps in the genetic algorithm and simultaneously exploits the learning with the actions that reduce the SPL by preventing the random actions letting more actions from optimal policy to learn optimality. This causes the initial SPL to rise and then gradually drop\cite{RLSurvey, adaptiveOpt}. The SPL keeps its decreasing tendency until t = 16, as the interventions are noticeably fewer as the DQN agent slowly enters the exploiting phase.

After t = 16, the SPL reaches the desired range of values (which gives the maximum reward of 10). More interventions are observed since now the jets have to maintain this level, instead of lowering it like before. However, fluctuations still occur. A reduction in effective SPL is observed.    

Looking at the velocity plot in \Cref{fig:stage_1}b, it can be seen that a velocity with a magnitude of around 1.0 can influence the flow to the extent we desire. Along with results from other trials, one of which will be presented in \Cref{fig:stage_1_test2}, it can be deduced that jet velocity with a magnitude above 3 shall likely cause simulation failure due to the breakdown of the PDE solver at high jet velocities due to the instability of the discretization scheme because of the high courant number near the jets. Essentially, it is a computational limitation and it can be tackled by using higher-order numerical schemes or finer discretization. In \Cref{fig:stage_1_test2}a, although the SPL is at the desired value, the simulation is stopped at just past t =10 (5000 mark on the x-axis). So, the lower value of the jet is at -3.5 and that is not ideal for the simulation. This threshold will be used to limit the jet speed in Stage 2 for the test cases.   

\subsection{Stage 2: Testing with definite jet velocity values}
\subsubsection{Set-up}
\begin{figure}[!t]
\centering
\includegraphics[width=0.48\textwidth]{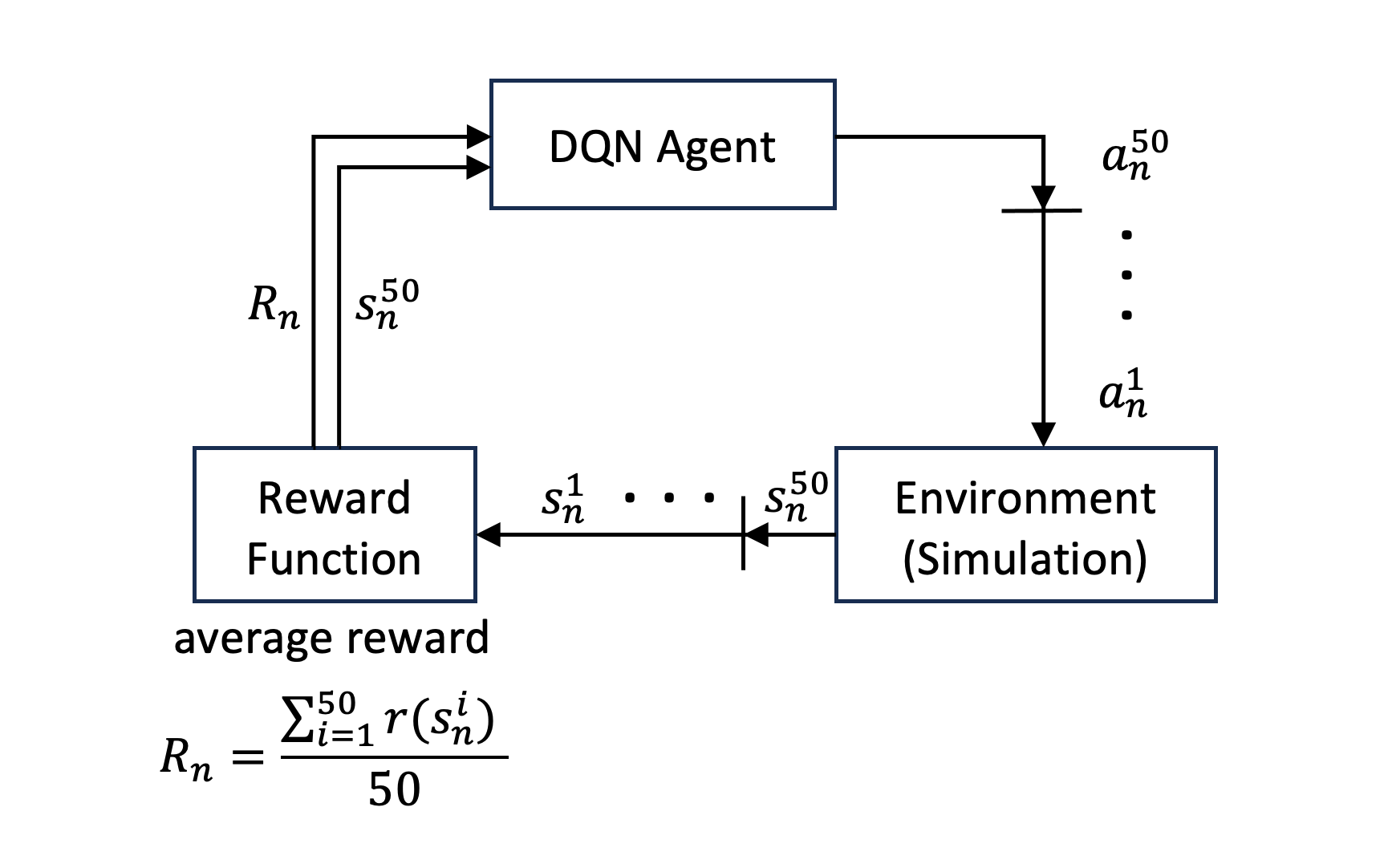}
\caption{Model of interaction between environment and DQN in Stage 2}
\label{fig:dqn_stg2}
\end{figure}
Based on the result of the previous section, we set a deterministic DQN algorithm with blowing and suction of three different values. In this stage, the jet velocity is no longer built up from the previous time step but rather has a distinct, fixed set of values. Two different test cases are presented: [±1.50, ±2.25, ±2.75] and [±2.00, ±2.50, ±3.00] (referred to as Test case 1 and Test case 2 respectively), to verify the result. The expectation from this stage is similar to the one before, which is lowering from 3 - 5 dB approximately. The reward calculation function is kept the same as in the previous stage (as in Table 1), but the reward calculation process is different. While the previous stage fully adopts the Markov model (\Cref{fig:RL_flow}), the strategy for this stage is accumulating the rewards between two interjections and averaging them, then returning to the DQN. Lastly, the 50-divisible state is returned to the DQN. Both test cases will implement this model. Refer to \Cref{fig:dqn_stg2} for further details.


\subsubsection{Results}
\begin{figure}[!t]
\centering
\includegraphics[width=0.485\textwidth]{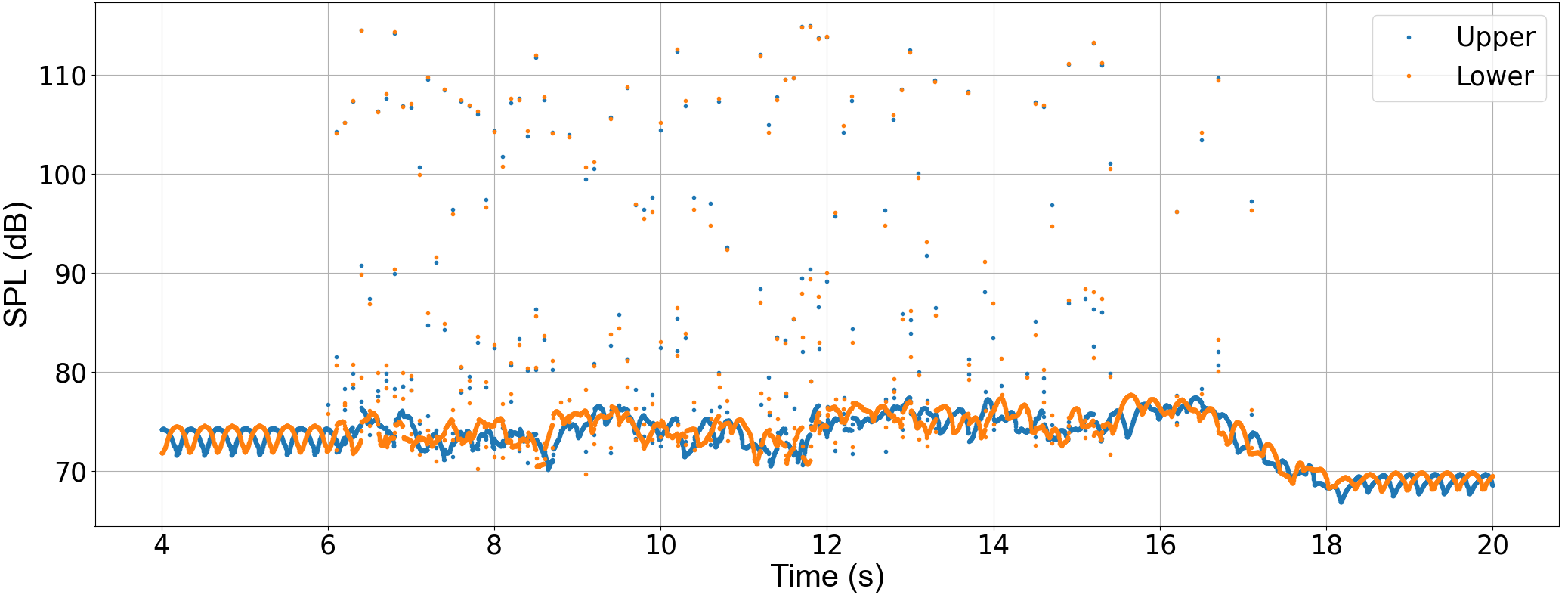}
\includegraphics[width=0.486\textwidth]{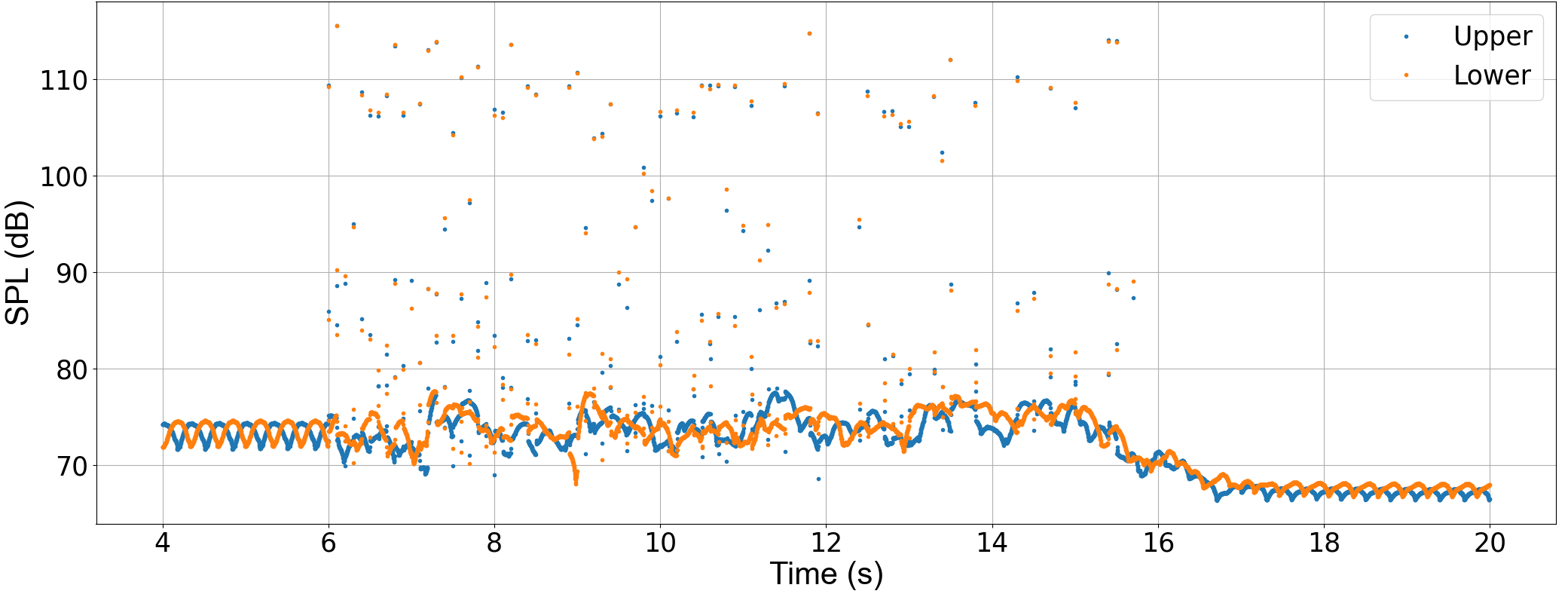}
\caption{a) Resulting SPL of Test case 1 ([±1.50, ±2.25, ±2.75]), b) Resulting SPL of Test case 2 ([±2.00, ±2.50, ±3.00])}
\label{fig:SPL_testcases}
\end{figure}

The resulting SPL of the upper and lower sensors are displayed in \Cref{fig:SPL_testcases} for both cases. Test case 1, using velocity values with lower magnitudes, converges to an SPL of just below 70 dB. Meanwhile, Test case 2 demonstrates a slightly lower SPL than that of Test case 1, and it seems to converge earlier as well. Also, the amplitude of oscillations is evident to have reduced.

\begin{figure}[!t]
\centering
\includegraphics[width=0.48\textwidth]{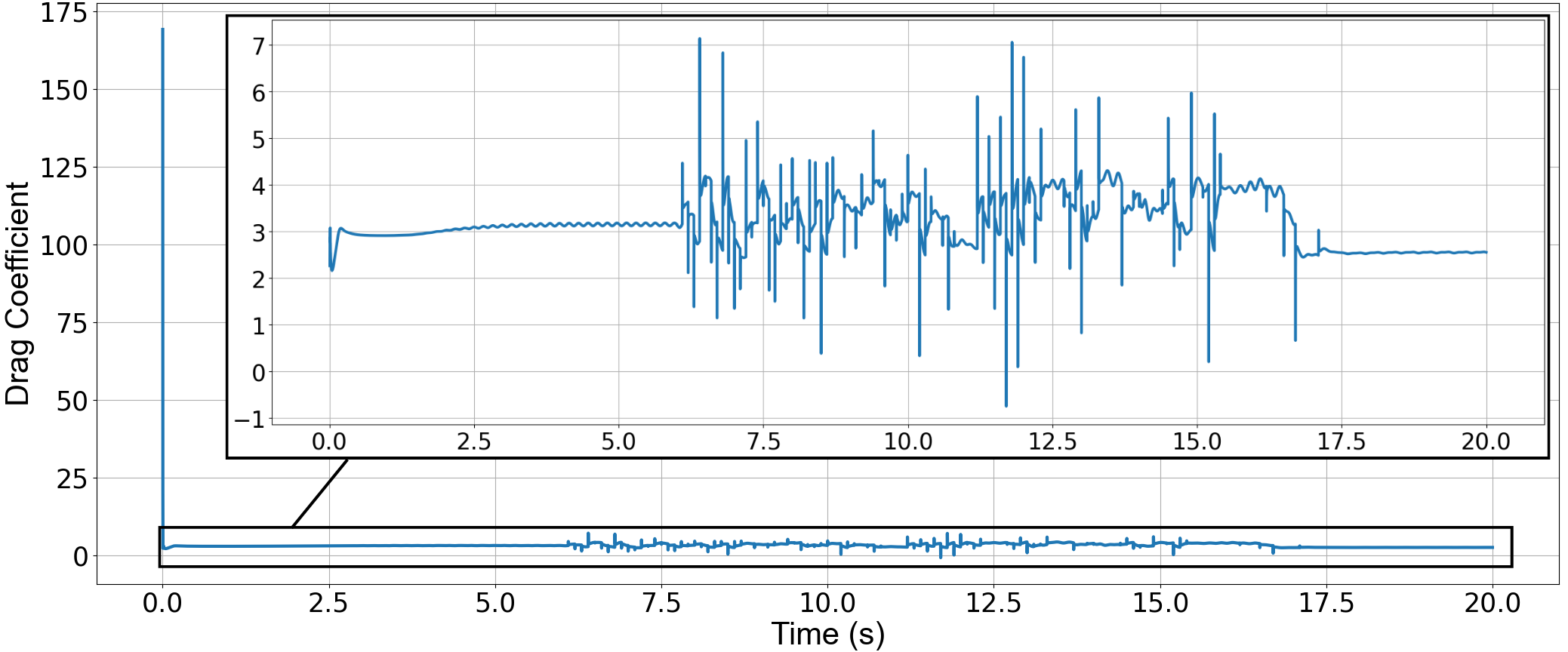}
\includegraphics[width=0.48\textwidth]{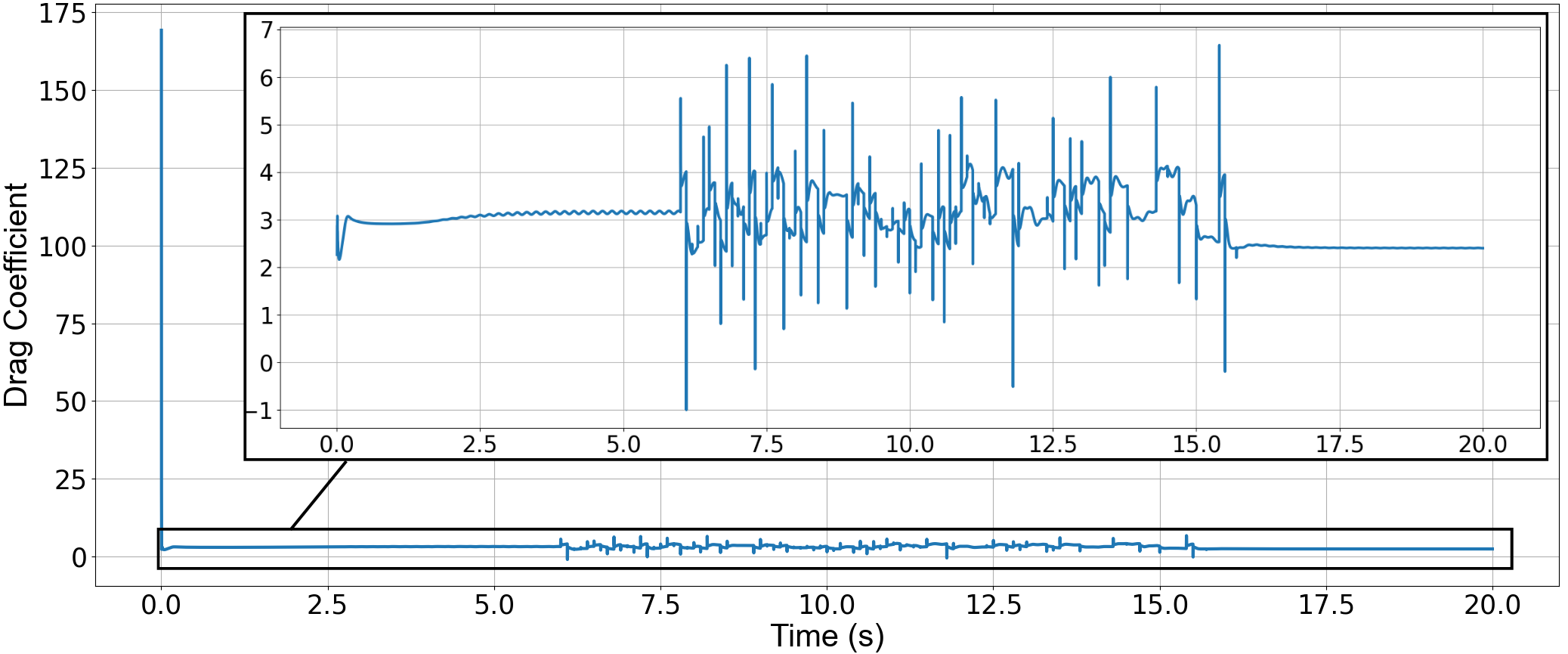}
\caption{Stage-2 a) Drag coefficient of Test case 1, b) Drag coefficient of Test case 2}
\label{fig:stage_2_drag}
\end{figure}

\begin{figure}[!t]
\centering
\includegraphics[width=0.48\textwidth]{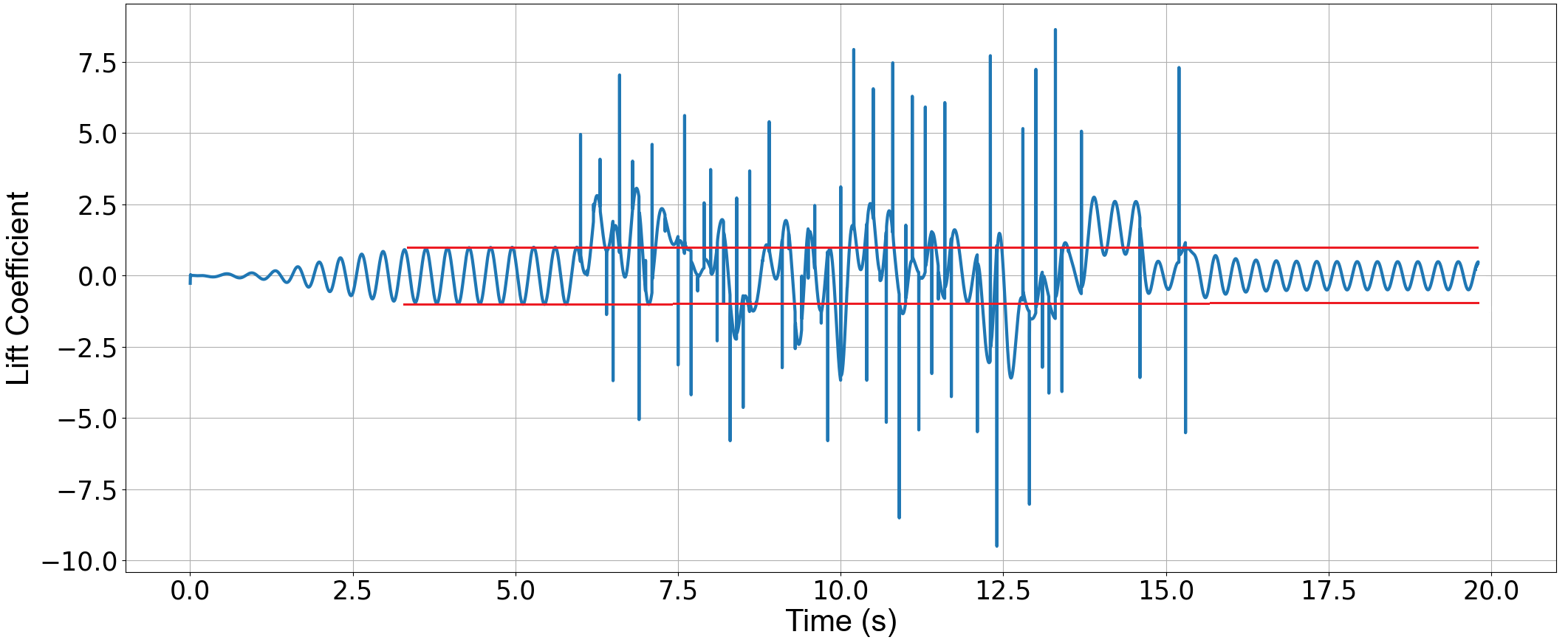}
\caption{Trend of lift coefficient during the process and notable reduction in the amplitude of oscillation}
\label{fig:stage_2_lift}
\end{figure}

\begin{figure}[!t]
\centering
\includegraphics[width=0.49\textwidth]{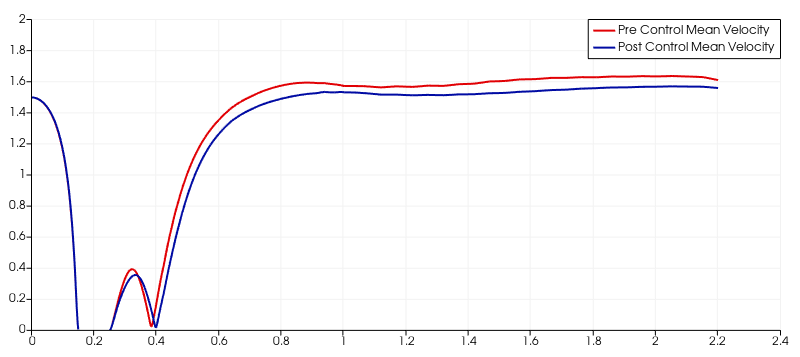}
\caption{Time averaged velocity without control (red) and after control (blue) along the longitudinal central line passing through the channel}
\label{fig:time_averaged_velocity_comparison}
\end{figure}

\begin{figure}[!t]
\centering
\includegraphics[width=0.525\textwidth]{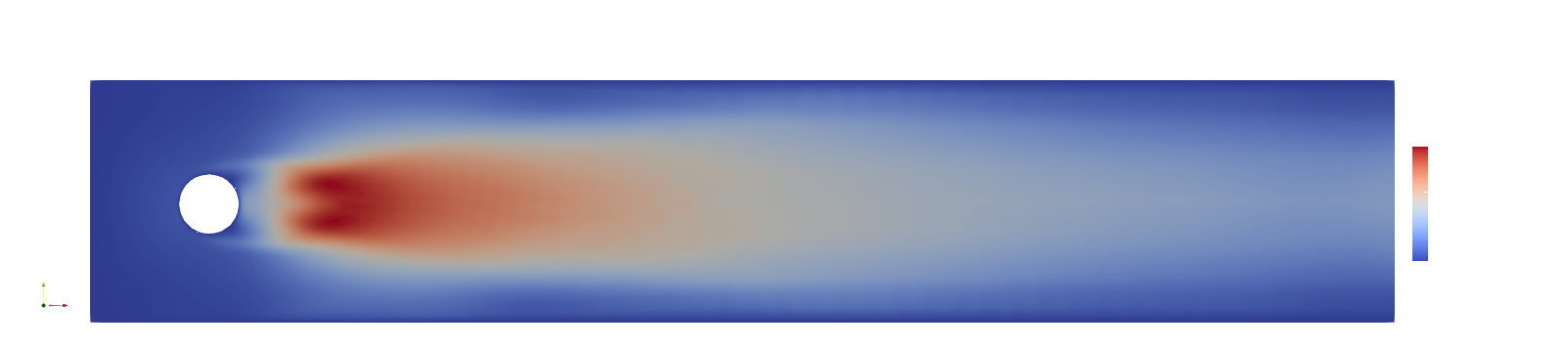}
\includegraphics[width=0.525\textwidth]{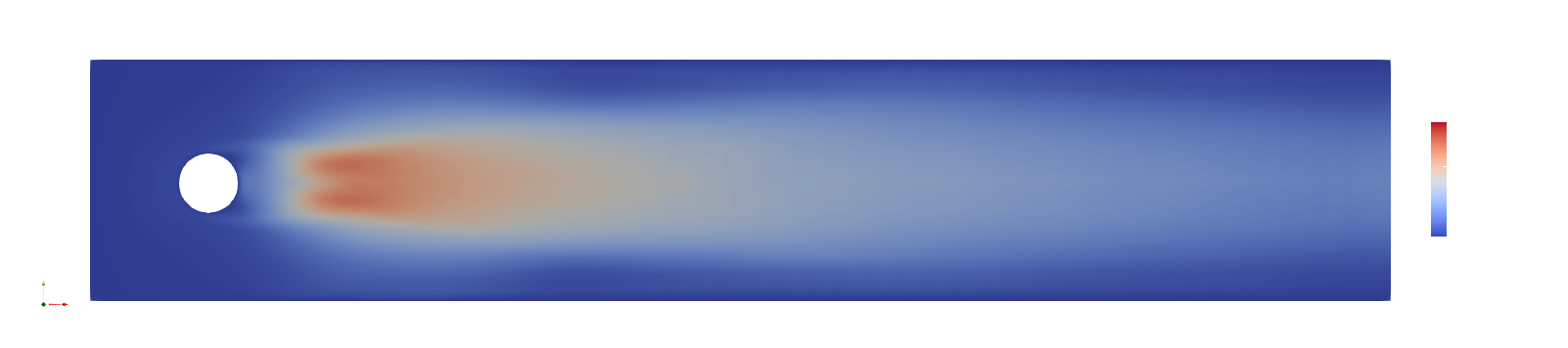}
\caption{Time averaged fluctuation without control (top) and after control (bottom)  }
\label{fig:time_averaged_fluctuation_comparison}
\end{figure}

\begin{figure}[!t]
\centering
\includegraphics[width=0.49\textwidth]{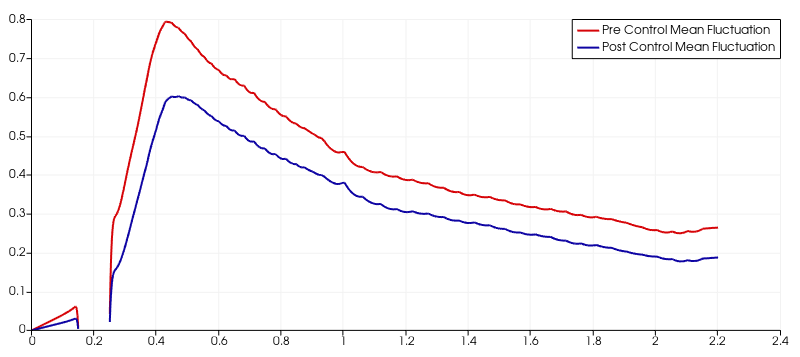}
\caption{Time averaged fluctuation without control (red) and after control (blue) along the longitudinal central line passing through the channel}
\label{fig:time_averaged_fluctuations_along_central_line}
\end{figure}

On the other hand, the instantaneous drag coefficient in each case is calculated and the result is impressive(refer to \Cref{fig:stage_2_drag}). The system experiences a large drag force at the start due to the direction of the flow, and this soon dissipates to lower values. The drag coefficient also expresses an oscillatory behaviour, although the amplitude is small. After great fluctuations in the exploration stage, the drag coefficient converges to a stable value, which is also lower than the initial drag and expresses oscillatory behaviour yet, with reduced amplitude. \Cref{fig:stage_2_lift} shows the reduction in the amplitude of oscillations in the lift force. When the time-averaged mean speed is calculated in the flow field along the longitudinal center line, the reduction after control is evident and shown in \Cref{fig:time_averaged_velocity_comparison}. A more important observation is the reduction in time-averaged standard deviation of flow speed along the downstream. The calculated field is a measure of averaged fluctuation shown in \Cref{fig:time_averaged_fluctuation_comparison}. A clear evidence of how the fluctuations are minimized due to the control is captured. The fluctuation is also measured along the same longitudinal center line to get a clearer idea of the magnitude of fluctuations before and after the control, which is \Cref{fig:time_averaged_fluctuations_along_central_line}.      


\subsection{Validation}
\subsubsection{Computational Framework}
Validation and numerical sensitivity analysis are necessary to establish the reliability of the computational framework. To do so, specifically, we conducted mesh independence analysis (\Cref{tab:mesh}) and time-step sensitivity analysis (\Cref{tab:timestep}).

Multiple mesh densities and time-step sizes were tested, and the resulting drag coefficient, and lift coefficient were monitored. The selected numerical parameters correspond to the regime where variations in the key flow quantities became negligibly small.

\begin{table}[!t]
\centering
\begin{tabular}{l|cccc}
\hline
\textbf{Mesh} & \textbf{Elements} & \textbf{$C_D$} & \textbf{$C_L$ (rms)}\\
\hline
Coarse & 42,000 & 3.08 & 0.212 \\
Medium & 85,000 & 3.14 & 0.226\\
Fine & 168,000 & 3.15 & 0.229\\
\hline
\end{tabular}
\caption{Mesh independence study for the uncontrolled cylinder flow case.}
\label{tab:mesh}
\end{table}

\begin{table}[!t]
\centering
\begin{tabular}{l|cccc}
\hline
\textbf{$\Delta t$} & \textbf{$C_D$} & \textbf{$C_L$ (rms)} & \textbf{Relative Error (\%)} \\
\hline
0.005 & 3.11 & 0.221 & 1.27 \\
0.002 & 3.14 & 0.226 & 0.31 \\
0.001 & 3.15 & 0.226 & -- \\
\hline
\end{tabular}
\caption{Time-step sensitivity analysis for the uncontrolled flow case.}
\label{tab:timestep}
\end{table}

\begin{table*}[!t]
\centering
\begin{tabular}{l|cc|cc|cc|cc}
\hline
\textbf{Case} & \textbf{$C_D$ (mean)} & \textbf{Red. (\%)} & \textbf{$C_D$ (std)} & \textbf{Red. (\%)} & \textbf{$C_L$ (std)} & \textbf{Red. (\%)} & \textbf{SPL (dB)} & \textbf{Red. (\%)} \\
\hline

Baseline (No Control) 
& 3.15 & 0
& 0.031 & 0
& 0.23  & 0
& 73.5  & 0 \\

DPM Adjoint (L1H100, \cite{liu2024adjoint}) 
& 2.35  & 25.4
& 0.0003 & 99.0
& 0.006 & 97.4
& -  & - \\

DRL using Velocity (L1H100, \cite{liu2024adjoint}) 
& 3.20 & - 1.6
& 0.024 & 22.6
& - & -
& - & - \\

DRL using Velocity (L2H512, \cite{liu2024adjoint}) 
& 3.03 & 3.8
& 0.014 & 54.8
& - & -
& - & - \\

DRL using SPL (\textbf{ours}, test case 1) 
& 2.65 & \textbf{15.9}
& 0.018 & \textbf{41.9}
& 0.15 & \textbf{34.8}
& 69.0 & \textbf{6.1}\\

DRL using SPL (\textbf{ours}, test case 2) 
& 2.40 & \textbf{23.8}
& 0.012 & \textbf{61.3}
& 0.11 & \textbf{52.2}
& 66.5 & \textbf{9.5} \\
\hline
\end{tabular}
\caption{Acoustic-based DRL-based flow control against baseline and benchmark data for flow past a circular cylinder.}
\label{tab:validation1}
\end{table*}

\subsubsection{Benchmarking}

\Cref{tab:validation1} further demonstrates that the acoustically informed DRL framework substantially outperforms velocity-based DRL configurations in terms of drag reduction and fluctuation suppression. This observation suggests that hydrodynamic pressure feedback provides a more physically informative control signal for wake stabilization under the present flow conditions. 

\subsubsection{Statistical Convergence}

Time averaging was performed only after the control policy reached a statistically stationary regime following the initial exploration phase. To verify the statistical convergence of the time-averaged flow quantities, convergence analyses of the mean velocity and vorticity fields were performed using progressively increasing averaging windows after the transient regime had decayed.
The time-averaged quantity was computed as:
\begin{equation}
\overline{\phi}=
\frac{1}{T}
\int_{t_0}^{t_0+T}
\phi(t)\,dt
\end{equation}

where $\phi$ denotes either the velocity or vorticity field, and $T$ represents the averaging duration.
To assess statistical convergence, the relative variation between successive averaging windows was evaluated as:

\begin{equation}
\epsilon(T)=
\frac{
\left\|
\overline{\phi}_{T+\Delta T}
-
\overline{\phi}_{T}
\right\|
}{
\left\|
\overline{\phi}_{T}
\right\|
}
\end{equation}

where $\|\cdot\|$ denotes the appropriate spatial norm. It was observed that both the mean velocity and vorticity fields reached statistically converged states with relative variations below $1\%$ after approximately 20 vortex shedding cycles.
Furthermore, the uncertainty associated with the averaged quantities was quantified using the temporal standard deviation:

\begin{equation}
\sigma_\phi=
\sqrt{
\frac{1}{N}
\sum_{i=1}^{N}
(\phi_i-\overline{\phi})^2
}
\end{equation}
and the standard error of the mean (SEM):

\begin{equation}
\mathrm{SEM}=
\frac{\sigma_\phi}{\sqrt{N}}
\end{equation}

where $N$ denotes the total number of statistically independent samples. The resulting uncertainty levels were found to be sufficiently small compared to the corresponding mean values, confirming the robustness and statistical reliability of the reported flow quantities. See the results in \Cref{tab:uncertainty}.

\begin{table}[!t]
\centering
\begin{tabular}{lccc}
\hline
\textbf{Quantity} & \textbf{Mean Value} & \textbf{Std. Dev.} & \textbf{SEM} \\
\hline
Streamwise Velocity $\overline{u}$ 
& 0.842 
& 0.021 
& 0.0021 \\
Vorticity Magnitude $\overline{\omega}$ 
& 5.37 
& 0.18 
& 0.018 \\
\hline
\end{tabular}
\caption{Statistical convergence and uncertainty analysis of time-averaged flow quantities.}
\label{tab:uncertainty}
\end{table}

\subsubsection{Sensor Placement}
Since the proposed control framework relies on pressure-based acoustic feedback, the spatial distribution and density of the sensor measurements can indeed influence the observed control behavior.

In the present study, the sensor surfaces were intentionally positioned around the wake region in order to capture the dominant hydrodynamic pressure fluctuations associated with vortex shedding. The upper and lower sensor arrays were selected specifically to monitor the alternating pressure signatures generated by the von Kármán vortex street.

The relatively dense sensor distribution was adopted to obtain a more accurate spatially averaged estimate of the wake-induced pressure fluctuations and reduce sensitivity to localized numerical oscillations.

To assess the robustness of the framework, additional tests were performed using reduced sensor densities and modified sensor locations. The overall control trends, including suppression of wake fluctuations and drag reduction, remained qualitatively consistent, although moderate quantitative variations in SPL reduction and convergence behavior were observed. See \Cref{tab:sensorplacement}.

These observations suggest that while sensor placement influences the sensitivity and responsiveness of the control policy, the proposed framework is not exclusively dependent on a single sensor configuration.
\begin{table}[!t]
\centering
\begin{tabular}{l|ccc}
\hline
\textbf{Sensor} & \textbf{SPL} & \textbf{Drag} & \textbf{Convergence} \\
\textbf{Configuration} & \textbf{Reduction (\%)} & \textbf{Reduction (\%)} & \textbf{Stability} \\
\hline

Coarse
& 
&  
& \\
(500 $\times$ 100)
& 7.8 
& 18.5
& Moderate \\
\hline
Medium
&  
&  
& \\
(1000$\times$250) 
& 8.9
& 21.3
& Good \\
\hline
Fine 
&  
&  
& \\
(2000 $\times$ 500) 
& 9.5 
& 23.8 
& Good \\
\hline
\end{tabular}
\caption{Sensitivity of control performance to sensor density.}
\label{tab:sensorplacement}
\end{table}






\section{Conclusion}
The results demonstrate that acoustically informed DRL-guided synthetic jet actuation can effectively attenuate wake fluctuations, reduce hydrodynamic pressure oscillations, and suppress drag-generating vortex structures in laminar cylinder wake flow. The framework further illustrates the feasibility of coupling reinforcement learning with pressure-based sensing for adaptive active flow control under nonlinear wake dynamics.

In this study, the sensibility of using noise as controlling parameter for flow control is suggested and discussed. Also, explored the application of Deep Reinforcement Learning (DRL) for active flow control to mitigate wake noise generated by a flow past a circular cylinder. Our approach involved employing hydrophone arrays or pressure sensors to capture acoustic signals and creating a feedback loop for a DRL agent to strategically control jet actuators placed on the cylinder's surface. The agent learned and adapted its control strategy based on the observed acoustic feedback, leading to a closed-loop control system. The results of our investigation demonstrated that DRL-based flow control effectively reduced wake intensity and the noise generated, and it also showed promising results in term of reducing drag. Not only the drag but also reduces the oscillations in drag and noise. This can play a crucial role in reduction of flutter in flow induced vibrations in marine oil rigs, aircraft wings etc. controlling hydrodynamic instabilities.

The study involved two main stages: the first stage aimed to explore the optimal range of jet velocity values and build a strategy for reducing noise. This stage revealed that jet velocities with a magnitude around 1.0 can influence the flow to achieve the desired SPL reduction. It also highlighted the importance of avoiding excessive jet velocities.

In the second stage, we conducted tests with fixed jet velocity values to verify the results from the exploration stage. The results showed that DRL-controlled jet actuators successfully achieved a significant reduction in SPL. Test case 2, using higher jet velocities, demonstrated comparatively better noise reduction and quicker convergence. The SPL without any control has a mean value of ~73.5dB. With the test case 1, flow control brings it down to ~69dB which is roughly 6.91\% reduction. Similarly, in test case 2 it reduces to ~66.5 which is a remarkable 9.5\% drop. Similarly, the drag coefficient without any control was oscillatory with a mean value of ~3.15. With the test case 1, flow control brings down the drag coefficient to ~2.65 which is roughly a remarkable 15.9\% reduction. Similarly, the coefficient in test case 2 sees a reduction of 23.8\% to clock a mean of ~2.4. The lateral oscillation due to lift forces is also remarkably dampened. Additionally, the study also observed that the drag coefficient experienced oscillatory behavior but converged to a stable trend, indicating that the DRL approach effectively controls the flow dynamics very much positively.

Although the present framework does not seek globally optimal control in the strict optimal-control sense, the DRL-guided strategy consistently identifies dynamically effective actuation regimes capable of reducing wake fluctuations, aerodynamic drag, and acoustic pressure levels. The results demonstrate the feasibility of acoustically informed reinforcement learning for adaptive wake stabilization under nonlinear flow conditions. This research underscores the potential of DRL algorithms with jet actuators, and sensor arrays as an add-on in active flow control. The findings open up new avenues for optimizing flow control in practical engineering applications and hold promise for reducing noise, drag, and enhancing the performance of various engineering systems. Future work in this area can explore more complex flow scenarios, further refine control strategies, and investigate the application of DRL in other engineering domains. The study serves as a stepping stone towards the integration of machine learning techniques for enhancing the efficiency and performance of active flow control systems.





\section*{Acknowledgement}
The authors wish to thank Prof. Rajeev Jaiman for his advice and help in proofreading the article. The authors would also like to thank Mr. Joseph Moster for his technology support. Thanks are also due to the Department of Mathematics at The University of British Columbia for granting access to infrastructural resources.
\section*{Declaration of competing interest}
The authors declare that they have no known competing financial
interests or personal relationships that could have appeared to influence
the work reported in this paper.
\section*{Data availability}
The authors declare that the data and code supporting the findings of this study are available within the paper and the GitHub repository: \href{https://github.com/Siddharth-Rout/FlowControlDRL}{github.com/Siddharth-Rout/FlowControlDRL}

\nocite{*}
\bibliography{asmejour}

\end{document}